\documentclass{article}
\usepackage{amssymb,theorem,amsmath}
\def\qed{{\hfill $\square$}}
\def\Me {\mathcal M}
\def\Ae {\mathcal A}
\def\Ha {\mathcal H}
\def\Be {\mathcal B}
\def\Fe {\mathcal F}
\def\Se {\mathcal S}

\def\Ka {\mathcal K}

\def\<{\langle}
\def\>{\rangle}
\def\supp{{\rm supp}\,}
\def\Tr {{\rm Tr}\,}
\newtheorem{thm}{Theorem}
\newtheorem{lemma}{Lemma}
\newtheorem{prop}{Proposition}
\newtheorem{coro}{Corollary}

\theorembodyfont{\rm}
\newtheorem{ex}{Example}
\newtheorem{rem}{Remark}

\begin{document}

\title{Reversibility conditions for quantum operations}
\author{Anna Jen\v cov\'a\\
{\small Mathematical Institute, Slovak Academy of Sciences,}\\
{\small \v {S}tef\'{a}nikova 49, 814 73 Bratislava, Slovakia} \\
{\small  jenca@mat.savba.sk}
}
\date{}

\maketitle

{\small \textbf{Abstract.} We give a list of  equivalent  conditions for
reversibility of the adjoint of a unital Schwarz map,
 with respect to a set of
quantum states. A large class of such conditions is given by preservation of
distinguishability measures: $f$-divergences, $L_1$ -distance, quantum Chernoff
and Hoeffding distances; here we summarize and extend the known results. Moreover,  we prove a number of conditions in terms of the properties of
a quantum Radon-Nikodym derivative  and  factorization  of states in the given set. Finally, we show that reversibility is equivalent to preservation of a large class of quantum Fisher informations and $\chi^2$-divergences. }

\section{Introduction}
In the mathematical description of quantum mechanics, a quantum mechanical system
 is represented by a $C^*$ - algebra $\Ae\subseteq B(\Ha)$ of bounded operators on a Hilbert space $\Ha$. In the case that $\Ha$ is finite dimensional, the physical states of the system are represented by density operators, that is, positive
operators with unit trace. The evolution of the system is described, in the
Schr\"odinger picture, by a transformation $T$ on the states. Here $T$ is usually
required to be a  completely positive trace preserving map on the  algebra.

Let $\Se$ be a set of states, then $\Se$ can be seen as carrying some
information. If $\Se$ undergoes a  quantum operation $T$, then some information
can be lost. If $\Se$ represents a code which is sent through a noisy channel $T:\Ae\to \Be$, then the resulting code $T(\Se)$ might contain less information than $\Se$. In the framework of quantum statistics, $\Se$ represents a prior knowledge
on the state of the system and the task of the statistician is to make some inference on the true state. But if, say, $\Se$ is a family of states on the bipartite
 system $\Ae\otimes \Be$ and only the  system $\Ae$ is accessible,
then the statistician has to work with the restricted states which might be
distinguished  with less precision. However, it might happen in some situations
 that the original information can be recovered, in the sense that there is a
quantum operation $S$ such that $S\circ T(\sigma)=\sigma$ for all $\sigma\in \Se$.
In this case we say that $T$ is reversible for $\Se$.
Such maps are also called  sufficient for $\Se$, which comes from the well known
notion of  sufficiency in classical statistics.

The information loss under quantum operations
is expressed in the monotonicity property of
 distinguishability measures: quantum $f$-divergences \cite{petz86b}
like relative entropy,
the $L_1$-distance, quantum Chernoff and Hoeffding distances \cite{auden}, etc.,
which means that these measures are nonincreasing under quantum operations.
It is quite clear that if $T$ is reversible for $\Se$, then $T$ must preserve
all of these measures on $\Se$.
It was an important observation in \cite{petz86} that preservation of the relative entropy, along with other  equivalent conditions,
 is equivalent to reversibility.  These results were then extended in the papers
 \cite{petz88,japetz,japetz2}, see also \cite{OP}.
The very recent paper \cite{bhmp} extends the monotonicity results to the case
 that $T$ is the adjoint of a subunital Schwarz map and proves that
reversibility  is equivalent to preservation of a large class of quantum
$f$-divergences, as well as distinguishability measures related to quantum hypothesis testing: the quantum Chernoff and Hoeffding distances. In the present paper, we find conditions for reversibility in terms of the $L_1$- distance
and  complete the results  for the Chernoff and Hoeffding distances and $L_1$- distance for $n$ copies of the states,  giving an answer to
some  of the questions left open  in \cite{bhmp}. Moreover, we find a class of quantum Fisher informations, such that
 preservation of elements in this class is equivalent to reversibility. We also prove reversibility conditions in terms   of a  quantum Radon-Nikodym derivative, and a quantum version
of the factorization theorem of classical statistics.

The various equivalent reversibility conditions are interesting also from the
opposite  point of view, when we are interested in the equality conditions for
the divergences in the first place. This was used, for example, for a
characterization of the quantum Markov property \cite{hjpw, japetz, ja10, jpp},
conditions for nullity of the  quantum discord \cite[Lemma 8.12]{hayashi06},\cite{datta}, conditions for strict decrease of Holevo quantity \cite{shirokov} and the equality conditions in certain
Minkowski type quantum inequalities and related quantities, \cite{jaru}.

 In a preliminary section, we deal with the properties of positive maps,
2-positive maps and Schwarz maps, and their duals with respect to a state.
In particular, we find a new  characterization of 2-positivity in terms of generalized Schwarz inequality and   we show that a unital positive  map has the property that its
duals with respect to all states are Schwarz maps, if and only if it is
 2-positive.
Then we  proceed to the various reversibility conditions:
we list  the  already known conditions related to  $f$-divergences
and give an example of a (non-quadratic and strictly convex) operator convex function $f$, such that
preservation of the corresponding $f$-divergence does not imply reversibility.
Further, we prove reversibility conditions in terms of
a quantum Radon-Nikodym derivative and
certain factorization  conditions on the states. In Sections 3.4 and 3.5 we deal
 with
 the $L_1$-distance, quantum Chernoff and Hoeffding distances.
In the last section, we give the reversibility conditions in terms of the quantum
Fisher information.

\section{Preliminaries}
Let  $\Ha$ be a finite dimensional Hilbert space and let
$\Ae\subseteq B(\Ha)$ be a $C^*$-algebra.  We denote by $\Ae^+$ the positive cone
in $\Ae$ and by $\Se(\Ae)$ the set of  states on $\Ae$. For $a\in \Ae^+$, we denote by $\supp a$ the projection onto the support of $a$, that is, $\supp a$ is the smallest projection $p$ satisfying $ap=a$.

A positive linear functional $\tau$ on $\Ae$ such that $\tau(ab)=\tau(ba)$ for all $a,b\in\Ae$ (equivalently,
$\tau(a^*a)=\tau(aa^*)$ for all $a\in \Ae$) is called a trace. We will also require that $\tau$ is faithful, then any linear functional  $\varphi$ on $\Ae$ has the form
 \[
\varphi(a)=\tau(a\rho_\varphi),\qquad a\in \Ae
 \]
for a unique operator $\rho_\varphi\in \Ae$, and  $\varphi$ is a state if and only if $\rho_\varphi\ge 0$ and $\tau(\rho_\varphi)=1$.
In this case, $\rho_\varphi$ is called the density operator of $\varphi$ with respect to $\tau$. Conversely, any operator $\rho\in \Ae^+$ with $\tau(\rho)=1$
 defines a state $\varphi_\rho$ on $\Ae$ with density $\rho$.
Moreover, if $\tau$ is faithful, then
\[
\<a,b\>_\tau =\tau (a^*b),\qquad a,b\in \Ae
\]
defines an inner product in $\Ae$.

 Clearly, $\Ae$  inherits the trace
$\Tr=\Tr_\Ha$ from $B(\Ha)$, but in
general, there exists different faithful traces on $\Ae$ even if we require $\tau(I)=\Tr(I)$. We will consider general traces  only in Section \ref{sec:fixed},
in the rest
  of the paper we always assume that $\tau=\Tr=\Tr_\Ha$ for a fixed representation $\Ae\subseteq B(\Ha)$.
Accordingly, the density operators with respect to $\Tr$ will be referred to simply as density operators and we  will identify $\Se(\Ae)$ with the set
$\{\rho\in\Ae^+, \Tr \rho=1\}$.
 We will also denote $\<a,b\>:=\<a,b\>_\Tr$ the restriction of
the Hilbert-Schmidt inner product in $B(\Ha)$.

\subsection{Positive maps}

Let $\Be\subseteq B(\Ka)$ be a finite dimensional $C^*$ algebra and let
$T: \Ae\to \Be$ be a positive  map. Let $T^*$ be the adjoint of $T$, with
respect to the Hilbert-Schmidt inner product.
We will say that $T$ is faithful if $T(a)=0$ for  $a\ge 0$  implies $a=0$.

\begin{lemma}\label{lemma:faith} Suppose that $T:\Ae\to \Be$ is a positive map. The following are equivalent.
\begin{enumerate}
\item[(i)]  $T(\rho)$ is invertible
for any positive invertible $\rho$.
\item[(ii)] $T(\rho)$ is invertible for some positive invertible $\rho$.
\item[(iii)]  $T^*$ is faithful.

\end{enumerate}
 \end{lemma}

{\it Proof.} The implication (i) $\implies$ (ii) is trivial.
Suppose (ii) and let $a\ge 0$ be
such that $T^*(a)=0$. Then
$0=\Tr T^*(a)\rho=\Tr a T(\rho)$, hence $a=0$.

Suppose (iii) and let $\rho$ be any positive invertible element.
Let $q:=\supp T(\rho)$. Then
$0=\Tr T(\rho)(I-q)=\Tr \rho T^*(I-q)$, this implies $I-q=0$, hence
 (i) holds.

\qed

\begin{lemma}\label{lemma:supp} Let $T:\Ae\to \Be$ be a positive map, such that
$T^*(I)\le I$.
Let $\rho$ and $\sigma$ be positive operators and let $p=\supp \rho$,
$p_0=\supp T(\rho)$,
$q=\supp \sigma$ and $q_0=\supp T(\sigma)$.
Then
\begin{enumerate}
\item[(i)] $T^*(I-p_0)\le I-p$
\item[(ii)] if $q\le p$ then $q_0\le p_0$
\item[(iii)] $T(p\Ae p)\subseteq p_0\Be p_0$.
\item[(iv)] if $T^*$ is unital, then $T^*(p_0)\ge p$.

\end{enumerate}

\end{lemma}

{\it Proof.} Note that for $0\le a\le I$ and any positive $\omega$,  $a\le I-\supp\omega$ if and only if
$\Tr a\omega=0$. We have
\[
\Tr \rho T^*(I-p_0)=\Tr T(\rho)(I-p_0)=0
\]
which implies (i). Moreover, suppose $q\le p$, then by (i),
\[
0\le\Tr T(\sigma)(I-p_0)=\Tr \sigma T^*(I-p_0)\le \Tr\sigma(I-p)=0
\]
this proves (ii). Let $a$ be a positive element in $p\Ae p$, then
$\supp a\le p$, hence by (ii), $\supp T(a)\le p_0$, so that $T(a)\in p_0\Be p_0$.
Since $p\Ae p$ is generated by its positive cone, this implies (iii).

Finally, (iv) follows directly from (i) if $T^*$ is unital.

\qed

 We say that $T$ is $n$-positive if the map
\[
T_{(n)}:= id_n \otimes T: M_n(\mathbb C)\otimes \Ae\to M_n(\mathbb C)\otimes \Be
\]
is positive, and $T$ is completely positive if it is $n$-positive for all
$n$.
The adjoint  $T^*$  is $n$-positive if and only if
$T$ is $n$-positive.

\subsection{2-positive maps and Schwarz maps}

We say that  $T$ is a Schwarz map if
it satisfies the Schwarz inequality
\begin{equation}\label{eq:schwarz}
T(a^*a)\ge T(a)^*T(a),\qquad a\in \Ae.
\end{equation}
This implies that $T$ is  positive and subunital, that is, $T(I)\le I$.
It is well known  that a unital 2-positive map is a Schwarz map \cite[Proposition 3.3]{paulsen}.

Let $c\in \Ae^+$ and $a\in \Ae$. We define $a^*c^{-1}a:=\lim_{\varepsilon \to 0}
a^*(c+\varepsilon I)^{-1}a$, if the limit exists.
Note that this is the case if and only if the  range of $a$ is contained in the
 range of $c$ and then $a^*c^{-1}a=ac^-a$, where $c^-$ denotes the generalized
inverse of $c$.

\begin{lemma}\label{lemma:bh} Let $a,b,c\in \Ae$. Then the block matrix
$M=\left(\begin{array}{cc} a & b\\
b^* &c \end{array}\right)$ is positive if and only if $c\ge 0$, $bc^{-1}b^*$
is defined and satisfies $a\ge bc^{-1}b^*$.

\end{lemma}

{\it Proof.} The proof for the case that  $c$ is invertible can be found in
\cite{bhatia}. For the general case, note that $M\ge 0$ if and only if
$\left(\begin{array}{cc} a & b\\
b^* &c+\varepsilon I \end{array}\right)$ is positive for all $\varepsilon >0$.
By the first part of the proof, this is equivalent to
$c\ge 0$ and $a\ge b(c+\varepsilon I)^{-1}b^*$ for all $\varepsilon >0$.
Since  $b(c+\varepsilon I)^{-1}b^*$ is an increasing net of positive operators,
 the limit exists if and only if it is bounded from above, this proves the Lemma.

\qed

Let $c\in \Ae$ be a positive invertible element. Then we say that $T$ satisfies
the  generalized Schwarz inequality  for $c$ if for all $a\in \Ae$,
$ T(a)^*T(c)^{-1}T(a)$ is defined and satisfies  \cite{liebru}
\begin{equation}\label{eq:rho_schwarz}
T(a^*c^{-1}a)\ge T(a)^*T(c)^{-1}T(a),\qquad a\in \Ae.
\end{equation}
Note that the condition that $T(a)^*T(c)^{-1}T(a)$ is defined is satisfied if
 $T^*$ is subunital, by Lemma \ref{lemma:supp} (iii).

The next proposition gives a characterization of 2-positivity of maps in terms of the generalized Schwarz inequality, which might be interesting in its own right:

\begin{prop}\label{prop:2pos} Let $T:\Ae\to \Be$ be a positive map.
Then $T$ is 2-positive if and only if $T$ satisfies the generalized Schwarz inequality for every positive invertible $c\in \Ae$.
\end{prop}

{\it Proof.} Let $M=\left(\begin{array}{cc} a & b\\
b^* &c \end{array}\right)$ be a positive element in $M_2(\Ae)$.  Let
$\varepsilon >0$ and denote $M_\varepsilon:=  \left(\begin{array}{cc} a & b\\
b^* &c+\varepsilon I \end{array}\right)$. Then  $M_\varepsilon \ge 0$ and it is clear that
$T_{(2)}(M)\ge 0$ if and only if $T_{(2)}(M_\varepsilon)\ge0$ for all
$\varepsilon >0$. Hence we may suppose that $c$ is invertible. In this case,
$M\ge 0$ if and only if
$ c\ge 0$  and $a-bc^{-1}b^*\ge 0$, by Lemma \ref{lemma:bh}. Then
\[
M=\left(\begin{array}{cc} a-bc^{-1}b^* & 0\\
0 & 0 \end{array}\right)+\left(\begin{array}{cc} bc^{-1}b^* & b\\
b^* &c\end{array}\right)
\]
where both summands are positive.  Since $T$ is positive, this implies  that
$T$ is 2-positive if and only if
 for all
$b\in \Ae$ and invertible $c\in \Ae^+$,
\[
T_{(2)}\left(\begin{array}{cc} bc^{-1}b^* & b\\
b^* &c\end{array}\right)=\left(\begin{array}{cc} T(bc^{-1}b^*) & T(b)\\
T(b^*) &T(c)\end{array}\right)\ge 0
\]
Again by Lemma \ref{lemma:bh}, this is equivalent to the generalized Schwarz inequality for $c$.

\qed

\subsection{The map $T_\rho$}

Let  $\rho\in \Se(\Ae)$.
We define a sesquilinear form  in $\Ae$ by
\[
\<a,b\>_\rho =\Tr a^*\rho^{1/2}b\rho^{1/2},\qquad a,b\in \Ae.
\]
Then $\<\cdot,\cdot\>_\rho$ defines an inner product in $p\Ae p$, where $p=\supp \rho$.

Let $T:\Ae\to \Be$ be a  positive and trace preserving map,
so that $T(\rho)$ is
a density operator in $\Be$.  Let $p_0=\supp T(\rho)$,  then by Lemma
 \ref{lemma:supp} (iii), $T(p\Ae p)\subseteq p_0 \Be p_0$.

The map  $T_\rho:p\Ae p\to  p_0\Be p_0$ is defined by
\[
T_\rho(b)= T(\rho)^{-1/2}T(\rho^{1/2}b\rho^{1/2})T(\rho)^{-1/2},\qquad b\in
 p\Ae p.
\]
Note that  $T_\rho(a)$ is  the unique element in $p_0 \Be p_0$
satisfying
\begin{equation}\label{eq:alpha_rho}
\<T^*(b),a\>_\rho=\<b,T_\rho(a)\>_{T(\rho)},\qquad b\in \Be,
\end{equation}
so that $T_\rho$ is the dual  of the unital map $T^*$, defined in \cite{petz84}.
Note also that  $T_\rho$ is positive and unital and its adjoint
$T^*_\rho: p_0\Be p_0\to p\Ae p$,
\[
T^*_\rho(b)=\rho^{1/2}T^*(T(\rho)^{-1/2}bT(\rho)^{-1/2})\rho^{1/2}
\]
satisfies
\begin{equation}\label{eq:alpha_rho_preserves}
T_\rho^*\circ T(\rho)=\rho
\end{equation}
by Lemma 2, (iv).

It can be shown that  $T$ is $n$-positive if and only if $T_\rho$ is $n$-positive.
We will now investigate the case when $T_\rho$ is a Schwarz map.

\begin{lemma} Let $T:\Ae \to \Be$ be a positive trace preserving map and suppose
that $\rho$ is an invertible density operator. Then
$T_\rho$ is a Schwarz map if and only if $T$ satisfies the generalized Schwarz
inequality for  $c=\rho$.

\end{lemma}

{\it Proof.} $T_\rho$ satisfies the Schwarz inequality (\ref{eq:schwarz}) if and
only if
\[
T(\rho^{1/2}b^*b\rho^{1/2})\ge T(\rho^{1/2}b^*\rho^{1/2})T(\rho)^{-1}T(\rho^{1/2}b\rho^{1/2}),\qquad b\in \Ae.
\]
Putting $a=\rho^{1/2}b\rho^{1/2}$, we see that this is equivalent to
\[
T(a^*\rho^{-1}a)\ge T(a)^*T(\rho)^{-1}T(a),\qquad a\in \Ae.
\]

\qed

 The above lemma, 
 together with Proposition \ref{prop:2pos}, implies the following result. Its
importance will  become clear at the beginning of Section \ref{sec:revers}.

\begin{prop}\label{prop:2_pos} Let $T:\Ae\to\Be$ be a positive trace preserving
map. Then
$T_\rho$ is a Schwarz map for any invertible density operator $\rho$ if and only
if $T$ is 2-positive.

\end{prop}

\subsection{Multiplicative domain and fixed points}\label{sec:fixed}

This  section contains some  known results on the multiplicative domains and
 sets of fixed points of unital Schwarz maps and related decompositions of
the density operators.
We include the proofs partly for the convenience of the reader,
 and partly because we need a particular form of some of the results (mainly
 Theorem \ref{thm:fixed_alpha_rho_alpha} (v) and (vi)) which might be difficult to find explicitly in the literature.

Let $\Be\subset \Ae\subseteq B(\Ha)$ be a $C^*$-subalgebra.
We will denote by $\Ae'$ the commutant of $\Ae$, that is the set of all elements
in $B(\Ha)$, commuting with $\Ae$.
Then $\Ae'$ is a $C^*$-subalgebra in $B(\Ha)$. The relative commutant of
$\Be$ in $\Ae$ is the subalgebra $\Be'\cap \Ae$.
A conditional expectation $E:\Ae\to \Be$ is a positive linear map, such that
$E(bac)=bE(a)c$ for all $a\in \Ae$, $b,c\in \Be$. Such a map is always completely positive. There exists
 a unique trace	 preserving conditional expectation $E: \Ae\to \Be$, determined by
 $\Tr(ab)=\Tr(E(a)b)$ for $a\in \Ae$, $b\in \Be$ (that is, $E$ is the adjoint
of the embedding $\Be\hookrightarrow \Ae$ with respect to $\<\cdot,\cdot\>$).

Let $\Phi:\Ae\to \Be$ be a unital Schwarz map. Let us denote
\[
\Me_{\Phi}:=\{a\in \Ae,\ \Phi(a^*a)=\Phi(a)^*\Phi(a), \ \Phi(aa^*)=
\Phi(a)\Phi(a)^*\}.
\]
It  is  known that \cite[Lemma 3.9]{bhmp}
\[
\Me_{\Phi}=\{a\in \Ae,\   \Phi(ab)=\Phi(a)\Phi(b),\
 \Phi(ba)=\Phi(b)\Phi(a), \forall b\in \Ae\}.
\]
This implies that  $\Me_\Phi$  is a subalgebra
 in $\Ae$, called the multiplicative domain of $\Phi$. The restriction of
$\Phi$ to  $\Me_\Phi$ is a *-homomorphism.

Let now $\Phi :\Ae\to \Ae$ be a unital Schwarz map and suppose that there is an
invertible density operator $\rho\in \Se(\Ae)$, such that $\Phi^*(\rho)=\rho$. Let
us denote
 by $\Fe_\Phi$ the set of fixed points of $\Phi$, that is,
\[
\Fe_{\Phi}:=\{a\in \Ae,\ \Phi(a)=a\}
\]
and let $\varphi_\rho$ denote  the state $\varphi_\rho(a)=\Tr \rho a$ for $a\in \Ae$. 

\begin{thm}\label{thm:fixed}
\begin{enumerate}
\item[(i)] $\Fe_\Phi$ is a subalgebra in $\Me_\Phi$.
\item[(ii)] There exists a conditional expectation $E_\Phi: \Ae\to \Fe_\Phi$,
such that $E_\Phi^*(\rho)=\rho$
\item[(iii)] $\rho^{it} \Fe_\Phi\rho^{-it}\subseteq \Fe_\Phi$ for all $t\in \mathbb R$.
\item[(iv)]  Let us fix a faithful trace $\tau$ in $\Fe_\Phi$.
Then we have a  decomposition
\[
\rho=\rho^A\rho^B,
\]
where $\rho^A\in \Fe_\Phi$ is the
density operator  with respect to $\tau$ of the restriction of  
$\varphi_\rho$ to $\Fe_\Phi$  and
$\rho^B\in \Fe_\Phi'\cap \Ae$ is a positive invertible element such that
$\Phi^*(\rho^B)=\rho^B$.
\end{enumerate}

\end{thm}

{\it Proof.}
(i) Let $a\in \Fe_\Phi$, then since $\Phi$ is a Schwarz map,
$\Phi(a^*a)\ge \Phi(a)^*\Phi(a)=a^*a$. But we have
$\Tr\rho(\Phi(a^*a)-a^*a)= 0$, so that $\Phi(a^*a)=a^*a$,  similarly
$\Phi(aa^*)=aa^*$, hence $a\in \Me_\Phi$. Let now $a,b\in \Fe_\Phi$, then
 $\Phi(ab)=\Phi(a)\Phi(b)=ab$ and obviously $\Phi(a+b)=a+b$, $\Phi(I)=I$,
 so that $\Fe_\Phi$ is a subalgebra.

(ii) Let $E_\Phi:= \lim_{n\to \infty} \tfrac 1n\sum_{k=0}^{n-1}\Phi^k$, then
by the ergodic theorem, $E_\Phi$  is  a conditional expectation
onto the fixed point subalgebra $\Fe_\Phi$. It is obvious that
$E_\Phi^*(\rho)=\rho$.

(iii) is equivalent to (ii) by Takesaki's theorem \cite{takesaki}.

(iv) It was shown in \cite{japetz} that for any subalgebra satisfying (iii), there is a
decomposition $\rho=\rho^A\rho^B$, where $\rho^A$ is the density of the
restriction of $\varphi_\rho$ to
$\Fe_\Phi$ with respect to $\tau$ and $\rho^B$ is a positive invertible element in the relative
commutant $\Fe_\Phi'\cap \Ae$. For any $a\in \Ae$,
\[
\Tr \Phi(a)\rho=\Tr \Phi(a) \rho^A\rho^B=\Tr \Phi(a\rho^A)\rho^B=
\Tr a\rho^A\Phi^*(\rho^B)
\]
so that $\rho^A\rho^B=\rho=\Phi^*(\rho)=\rho^A\Phi^*(\rho^B)$, this implies
$\rho^B=\Phi^*(\rho^B)$.

\qed

\begin{thm}\label{thm:fixed_alpha_rho_alpha} Let
$\rho\in \Ae$ be an invertible  density operator and let $T:\Ae\to \Be$ be a trace
preserving map, such that both $T^*$ and $T_\rho$ are Schwarz maps and
$\rho_0:=T(\rho)$ is invertible.
Denote $\Phi:= T^*\circ T_\rho$ and $\tilde \Phi:=
T_\rho\circ T^*$.
Then
\begin{enumerate}
\item[(i)] $\Fe_{\tilde \Phi}$ is a subalgebra in $\Me_{T^*}$
and $\Fe_{\Phi}$ is a subalgebra in $\Me_{T_\rho}$.
\item[(ii)] The restriction of $T^*$ is a  *-isomorphism from
$\Fe_{\tilde \Phi}$ onto $\Fe_{\Phi}$, and its inverse is the
restriction of $T_\rho$.
\item[(iii)] $\Fe_{\Phi}$ is a subalgebra in $T^* (\Me_{T^*})$.
\item[(iv)] $\rho^{it}\Fe_{\Phi}\rho^{-it}\subseteq
\Fe_{\Phi}$ and $\rho_0^{it}\Fe_{\tilde\Phi}\rho_0^{-it}\subseteq
\Fe_{\tilde\Phi}$, for all $t\in \mathbb R$.
\item[(v)] $T(\Fe_{\Phi}'\cap\Ae)\subseteq \Fe_{\tilde \Phi}'\cap \Be$
\item[(vi)] There are  decompositions
\[
\rho=T^*(\rho_0^A)\rho^B,\qquad \rho_0= \rho^A_0 T(\rho^B)
\]
where $\rho^A_0\in \Fe_{\tilde \Phi}^+$
 and $\rho^B\in \Fe_{\Phi}'\cap \Ae^+$ is such that
$\Phi^*(\rho^B)=\rho^B$.
 \end{enumerate}

\end{thm}

{\it Proof.} Note that we have $\Phi^*(\rho)=\rho$ and
$\tilde \Phi^*(\rho_0)=T\circ\Phi^*(\rho)=
\rho_0$. Moreover, since $T_\rho^*(\rho_0)=\rho$, $T_\rho$ is faithful
by Lemma \ref{lemma:faith}.

 By Theorem \ref{thm:fixed} (i), $\Fe_{\tilde \Phi}$ is a subalgebra
in $\Me_{\tilde \Phi}$. It is easy to see that, since $T_\rho$ is
faithful,  $\Me_{\tilde \Phi}\subseteq \Me_{T^*}$. The second inclusion
in  (i) is proved similarly.

 By (i), the restriction of $T^*$ is a
*-homomorphism on
$\Fe_{\tilde \Phi}$. Since
$\Phi\circ T^*=T^*\circ \tilde \Phi$ and
$\tilde\Phi\circ T_\rho=T_\rho\circ \Phi$,
we have $T^*(\Fe_{\tilde \Phi})\subseteq \Fe_{\Phi}$,
$T_\rho(\Fe_{\Phi})\subseteq \Fe_{\tilde \Phi}$ and
$T_\rho\circ T^*(a)=a$ for  $a\in \Fe_{\tilde\Phi}$, this proves
(ii).

(iii) follows from (i) and (ii).

(iv) follows from Theorem \ref{thm:fixed} (iii).

To prove (v), let $b\in \Fe_{\tilde \Phi}$, $a\in \Fe_{\Phi}'\cap \Ae$ and $c\in
\Be$. Then
\begin{eqnarray*}
\Tr c b T(a)&=&\Tr T^*(cb)a=\Tr T^*(c)T^*(b)a=
\Tr T^*(c)aT^*(b)=\Tr T^*(b)T^*(c)a\\
&=&
\Tr T^*(bc)a=\Tr bcT(a)=\Tr cT(a)b
\end{eqnarray*}
so that $T(a)\in \Fe_{\tilde\Phi}'\cap \Be$, where we used
 the fact that $b\in \Me_{T^*}$,
 $T^*(b)\in \Fe_\Phi$ and cyclicity of the trace.

To prove (vi), let $\tau$ be the restriction of $\Tr$ to  $\Fe_{\Phi}$.
By (ii), $\tilde \tau:=\tau\circ T^*$
 defines a faithful trace on $\Fe_{\tilde\Phi}$.
  By  Theorem \ref{thm:fixed} (iv), we have the decompositions
\[
\rho=\rho^A\rho^B,\qquad \rho_0=\rho_0^A\rho_0^B
\]
where $\rho^A$ ($\rho_0^A$) is the density of the
restriction of $\varphi_\rho$ ($\varphi_{\rho_0}$) to
$\Fe_{\Phi}$ ($\Fe_{\tilde \Phi}$) with respect to $\tau$ ($\tilde \tau$).
Let now $a\in \Fe_{\Phi}$, then
\begin{eqnarray*}
\tau (a T^*(\rho^A_0))&=& \tau(\Phi(a) T^* (\rho^A_0))=
\tau( T^*(T_\rho(a))T^*(\rho^A_0))=
\tau( T^*(T_\rho(a)\rho^A_0)) \\
&=& \tilde \tau( T_\rho(a)\rho^A_0)
= \Tr T_\rho(a)\rho_0=\Tr a\rho=\tau(a\rho^A)
\end{eqnarray*}
It follows  that $\rho^A=T^*(\rho^A_0)$. If $b\in \Be$, then
\[
\Tr T^*(b)\rho =\Tr T^*( b)T^*(\rho^A_0)\rho^B=
\Tr T^*(b\rho^A_0)\rho^B=\Tr b\rho^A_0 T(\rho^B)
\]
so that $\rho_0=\rho_0^AT(\rho^B)$.

\qed

\section{Conditions for reversibility}\label{sec:revers}

Let $\Ae\subseteq B(\Ha)$ and $\Be\subseteq B(\Ka)$ be finite dimensional
$C^*$-algebras.
Let $\Se\subset \Se(\Ae)$ be a  set of density operators and let
$T: \Ae\to \Be$ be such that $T^*$ is a unital Schwarz map. We say that $T$
is reversible (or sufficient) for $\Se$ if there is a  map $S:\Be\to \Ae$,
such that $S^*$ is a unital Schwarz map and
\begin{equation}\label{eq:rev}
S\circ T(\sigma)=\sigma,\qquad \sigma\in \Se.
\end{equation}
In this section, we study various conditions for reversibility.
If not stated otherwise, we  assume that the following two conditions hold:

\begin{enumerate}

\item $\Se$ contains an invertible element $\rho$ and
$T(\rho)$ is invertible as well.
\item $T:\Ae \to \Be$ is such that both $T^*$ and $T_\rho$ are unital Schwarz maps.
 \end{enumerate}

In the original approach of \cite{petz88}, the map $T$ and the recovery map $S$ 
were both  required to be 2-positive.
The possibility of weakening this assumption was discussed in 
\cite[Remark 5.8]{bhmp}, where the question was raised whether it is enough to 
assume that $T^*$ is a unital Schwarz map for the map $T_\rho$ to be a Schwarz map as well. Proposition \ref{prop:2_pos} above shows that this is not the case, in 
fact, it follows that  if condition 2 holds for any density $\rho$, then  $T$ 
must be 2-positive. Moreover, as we will see in
Theorem \ref{thm:rev_d},  regarding reversibility of $T$, the condition 
2 is not more general than assuming
 that $T$ is a completely positive map.

On the other hand, note  that the condition  1 is not restrictive. 
Indeed, for  $\Se\subset \Se(\Ae)$
  there always
exists a (finite) convex combination $\rho$ of elements in $\Se$, such that
$\supp \sigma \le\supp \rho=:p$ for  all $\sigma\in \Se$. Moreover,
$T$ is reversible for $\Se$ if and only if it is reversible for the  closed
convex hull
$\bar{co}(\Se)$, therefore, we may always  suppose that $\rho\in \Se$. 
By Lemma \ref{lemma:supp}, we also have $p_0:=\supp T(\rho) \ge \supp T(\sigma)$
 for all $\sigma\in \Se$. 
Hence $\Se\subset \Se(p\Ae p)$ and $T(\Se)\subset \Se(p_0\Be p_0)$.

Let $\tilde T$ be the restriction of $T$ to $p\Ae p$, then $\tilde T$ maps 
$p\Ae p$ into $p_0\Be p_0$, by Lemma \ref{lemma:supp}.
We have  $\tilde T(\sigma)=T(\sigma)$ for $\sigma\in \Se$.  Again by Lemma 
\ref{lemma:supp}, 
\[
\tilde T^*(p_0)=pT^*(p_0)p=p,
\]
 so that $\tilde T^*$ is a unital 
Schwarz map. Note also that $\tilde T_\rho=T_\rho$.
It follows that if $T$ satisfies the condition 2, then $\tilde T$ satisfies 
both 1 and 2.
Moreover, $T$ is reversible for $\Se\subset \Se(\Ae)$ if and only if 
$\tilde T$ is reversible for $\Se\subset \Se(p\Ae p)$. Indeed, 
let $\tilde S$ be the restriction of $S$ to $p_0\Be  p_0$, where $S:\Be\to \Ae$ 
is the adjoint of  a unital 
Schwarz map satisfying (\ref{eq:rev}). 
Then $\tilde S$ maps $p_0\Be p_0$ into $p\Ae p$, $\tilde S^*$
is a unital Schwarz map and $\tilde S\circ \tilde T(\sigma)=S\circ T(\sigma)=\sigma$ for all $\sigma\in \Se$. Conversely, let
 $\tilde S: p_0 \Be p_0\to p\Ae p$ be the adjoint of  a  unital Schwarz map, 
such that 
$\tilde S\circ \tilde T(\sigma)=\sigma$ for $\sigma\in \Se$, then we extend 
$\tilde S$ to a map   $S:\Be\to \Ae$ by
\[
S(b)= \tilde S(p_0bp_0)+ [\Tr b(1-p_0)]\rho  \qquad b\in \Be 
\]
Then $S^*$  is a unital Schwarz map and $S\circ T(\sigma)=\tilde S\circ \tilde 
T(\sigma)
=\sigma$ for every $\sigma\in \Se$. Moreover, $S$ is $n$-positive whenever $\tilde S$ is $n$-positive.

The above constructions can be easily illustrated in the trivial case when $\Se=\{\rho\}$. Then both $T$ and $\tilde T$ are always reversible, the recovery map 
being  $T_\rho^*$ for $\tilde T$, and  an extension of $T_\rho^*$ for $T$.

\subsection{Quantum $f$-divergences}

Let $f:[0,\infty)\to \mathbb R$ be a function. Recall that $f$ is operator convex if $f(\lambda A+(1-\lambda)B)\le \lambda f(A)+
(1-\lambda)f(B)$ for any $\lambda\in [0,1]$ and any positive matrices $A,B$ of any dimension. It was proved in \cite{bhmp} that any operator convex function
 has an integral representation of the form
 \[
f(x)=f(0)+ax+bx^2+\int_{(0,\infty)}(\frac{x}{1+t}-\frac x{x+t})d\mu_f(t),\qquad x\in [0,\infty)
 \]
where $a\in \mathbb R$, $b\ge 0$ and $\mu_f$ is a non-negative measure on $(0,\infty)$ satisfying $\int (1+t)^{-2}d\mu_f(t)<\infty$.

Let now $\sigma$ and $\rho$ be two density operators and suppose that $\supp\sigma\le \supp \rho$. Let $\Delta_{\sigma,\rho}=L_\sigma
R^{-1}_\rho$ be  the relative modular operator, note that $\Delta_{\sigma,\rho}(a)=\sigma a\rho^{-1}$ for any $a\in \Ae$. Let $f:[0,\infty)\to \mathbb R$ be
an  operator convex function. The
$f$-divergence of $\sigma$ with respect to $\rho$ is defined by
\[
S_f(\sigma,\rho)=\< \rho^{1/2}, f(\Delta_{\sigma,\rho})\rho^{1/2}\>
\]
see \cite{bhmp} also for the case of arbitrary pairs of density operators.
A well-known example is the relative entropy $S(\sigma,\rho)=\Tr \sigma(\log\sigma-\log\rho)$, which corresponds to the
operator convex function $f(x)=x\log x$. Another example is given by
$S_s(\sigma,\rho)=1-\Tr\sigma^s\rho^{1-s}$, this corresponds to the function $f_s(x)=1-x^s$, which is operator convex for $s\in [0,1]$.

Let $T^*:\Be\to \Ae$ be a unital Schwarz map. Then any $f$-divergence is monotone under $T$ \cite{bhmp},
in the sense that
\[
S_f(T(\sigma),T(\rho))\le S_f(\sigma,\rho).
\]

\begin{thm}\label{thm:diverg}\cite{bhmp} Under the conditions 1 and 2,
 the following are equivalent.
\begin{enumerate}
\item[(i)] $T$ is reversible for $\Se$.
\item[(ii)] $S(T(\sigma),T(\rho))=S(\sigma,\rho)$ for all $\sigma\in \Se$.
\item[(iii)] $T^*(T(\sigma)^{it}T(\rho)^{-it})=\sigma^{it}\rho^{-it}$ for
 $\sigma\in \Se$,  $t\in \mathbb R$.
\item[(iv)] $\Tr T(\sigma)^sT(\rho)^{1-s}=\Tr \sigma^s\rho^{1-s}$  for all
$\sigma\in \Se$ and some $s\in (0,1)$.
\item [(v)] $S_f(T(\sigma),T(\rho))=S_f(\sigma,\rho)$ for all $\sigma\in \Se$ and some operator convex function $f$ with
$|\supp \mu_f|\ge \dim(\Ha)^2+\dim(\Ka)^2$, where  $|X|$ denotes
 the number of elements in the set $X$.
\item[(vi)] Equality holds in (v) for all operator convex functions.
\item[(vii)] Equality holds in (iv) for all $s\in [0,1]$.
\item[(viii)] $T_\rho^*\circ T(\sigma)=\sigma$ for all $\sigma\in \Se$.

\end{enumerate}
\end{thm}

\begin{rem} The equivalence of (i)-(iii) and (viii) was first proved in
\cite{petz86}, for the case when all states are faithful and
$T$ is the restriction to a subalgebra, and
subsequently for any unital 2-positive map in \cite{petz88}, in the more general
setting of von Neumann algebras, see also \cite{japetz} and \cite{japetz2}, where conditions (iv), (vii) were proved.

\end{rem}

The following example shows that, unlike the classical case, preservation of an
$f$-divergence with strictly operator convex $f$ is in general not sufficient for reversibility. This
 solves another open problem of \cite{bhmp}, showing that the support condition
in Theorem \ref{thm:diverg} (v) cannot be completely removed.

\begin{ex} The function  $f(x)=(1+x)^{-1}$, $x\ge 0$
 is operator convex and the corresponding measure $\mu_f$ is concentrated in the point $t=1$, $\mu_f(\{1\})=1$. We have
\[
S_f(\sigma,\rho)=\Tr \rho(L_\sigma+R_\rho)^{-1}(\rho)
\]
We will show that the equality
$S_f(T(\sigma),T(\rho))=S_f(\sigma,\rho)$ does not imply reversibility of $T$.

Let $\Ae$ be a matrix algebra and let $\sigma\in \Ae$ be an invertible
density matrix. Let  $p\in \Ae$ be a projection such that $\sigma p\ne p\sigma$
 and $\Tr p\sigma=\lambda\ne 1/2$. Let $\Be\subset\Ae$ be the abelian
subalgebra generated  by $p$ and let $T:\Ae\to \Be$ be the trace preserving
conditional  expectation, then $T(\sigma)$ is the density of the restriction
 of $\sigma$ to $\Be$.  Put $x:=(1-\lambda)p+\lambda(I-p)\in \Be$ and
$\rho:=(I-x)^{-1}\sigma x$. Then
\[
\rho =c^{-1}x\sigma x\ge 0
\]
where $c=\lambda(1-\lambda)$, and
\[
\Tr \rho=c^{-1} \Tr \sigma x^2=1
\]
so that $\rho$ is an invertible  density matrix as well. Moreover, we also have
$T(\rho)=(I-x)^{-1}T(\sigma)x$, so that
\[
(L_\sigma+R_\rho)^{-1}(\rho)=x=(L_{T(\sigma)}+R_{T(\rho)})^{-1}(T(\rho))
\]
and the equality $S_f(T(\sigma),T(\rho))=S_f(\sigma,\rho)$ holds. On the other
 hand, we have from Theorem \ref{thm:factor} (iv) below that  $T$ is reversible if and
only if $\sigma=\sigma^A\rho^B$ and $\rho=\rho^A\rho^B$ for some
$\sigma^A,\rho^A\in \Be^+$ and $\rho^B\in \Ae^+$. It follows that both $\sigma^A$ and $\rho^A$ commute with $\rho^B$ and, since $\Be$ is abelian, this implies that $\sigma^A$ and $\rho^A$ commute with $\sigma$. But this is
possible only if $\rho^A$ and $\sigma^A$ are constants. It follows that we must have $\sigma=\rho$ and it is easy to see that this implies that $\sigma$ commutes with $x$, which is not possible by the construction of $x$.

\end{ex}

\subsection{The commutant Radon-Nikodym derivative}

Let $\rho$, $\sigma$ be density operators in $\Ae$ and suppose that
$\supp \sigma\le \supp \rho=:p$. The commutant
 Radon-Nikodym derivative of $\sigma$ with respect to $\rho$ is defined by
\[
d(\sigma,\rho)=\rho^{-1/2}\sigma\rho^{-1/2}
\]
Then $d=d(\sigma,\rho)$ is the unique element in $p\Ae p$,
satisfying
\begin{equation}\label{eq:drs}
\Tr \sigma a=\<I,a\>_{\sigma}=\<d,a\>_{\rho}
\end{equation}
Moreover, $d\ge 0$ and $\|d\|$ is the smallest number $\lambda$ satisfying
$\sigma \le \lambda \rho$, note that $\|d\|\ge 1$ and $\|d\|=1$ if and only if $\rho =\sigma$.

\begin{lemma}\label{lemma:d_0} Let $\supp\sigma\le \supp\rho$.
Let  $T:\Ae\to \Be$ be  a trace preserving  positive  map. Then
\[
d(T(\sigma),T(\rho))=T_\rho(d(\sigma,\rho))
\]
\end{lemma}

{\it Proof.} Directly by definition of $T_\rho$ and $d(\sigma,\rho)$.

\qed

The following simple  lemma provides a useful tool for the analysis of
reversibility. Note also that it gives a reversibility condition also for the case when   both
$T$ and the reverse map  $S$  are only required to be positive
and  trace preserving.

\begin{lemma}\label{lemma:alpha_d} Let $\rho$ be invertible and let $T:\Ae\to \Be$ be
 a trace preserving positive map. Then
$T_\rho^*\circ T(\sigma)=\sigma$ if and only if $T^*(d(T(\sigma),T(\rho)))=d(\sigma,\rho)$.
\end{lemma}

{\it Proof.}
For $a\in \Ae$, we have by (\ref{eq:alpha_rho}) and (\ref{eq:drs}) that
\begin{eqnarray*}
\<T^*(d(T(\sigma),T(\rho))),a\>_\rho&=&
\<d(T(\sigma),T(\rho)),T_\rho(a)\>_{T(\rho)}=\Tr T_\rho(a)T(\sigma)\\
&=&
\Tr a T_\rho^*\circ T(\sigma)
\end{eqnarray*}
It follows that $T^*(d(T(\sigma),T(\rho)))=d(\sigma,\rho)$ if and only if
$\Tr a T_\rho^*\circ T(\sigma)=\Tr a\sigma$ for all $a\in \Ae$.

\qed

Now we are able to characterize reversibility in terms of the Radon-Nikodym 
derivative. While (ii) or (iii) give easy conditions for reversibility, the 
condition (iv) will be necessary for the proof of Theorem \ref{thm:suff2_d} below. The last two conditions are not really new, but will be useful in proving Theorem \ref{thm:qcbhb}.

\begin{thm}\label{thm:rev_d} Suppose the conditions 1 and 2 hold. Let us denote
$\Phi=T^*\circ T_\rho$.
Then the following are equivalent.

\begin{enumerate}
\item[(i)] $T$ is reversible for  $\Se$.
\item[(ii)]$T^*(d(T(\sigma),T(\rho)))=d(\sigma,\rho)$, for all $\sigma\in \Se$.
\item[(iii)]  $d(\sigma,\rho)\in \Fe_{\Phi}$, for all $\sigma\in \Se$.
\item[(iv)]  $\rho^{it} d(\sigma,\rho)\rho^{-it}\in T^*(\Me_{T^*})$,
for all $\sigma\in \Se$ and $t\in \mathbb R$.
\item[(v)] There is a trace preserving  completely positive map $\hat S:\Be\to \Ae$, such that
$\hat S\circ T(\sigma)=\sigma$, $\sigma\in \Se$.
\item[(vi)] There are trace preserving  completely positive maps $\hat T:\Ae\to \Be$
 and $\hat S:\Be\to \Ae$, such that $\hat T(\sigma)=T(\sigma)$,
$\hat S\circ T(\sigma)=\sigma$, $\sigma\in \Se$.
\end{enumerate}

\end{thm}

{\it Proof.} By Lemma \ref{lemma:alpha_d}, (ii) is equivalent to
$T_\rho^*\circ T(\sigma)=\sigma$ for $\sigma\in \Se$, which is equivalent to (i)
by Theorem \ref{thm:diverg} (viii).
(iii) is the same as (ii), by Lemma \ref{lemma:d_0}. Since by
Theorem \ref{thm:fixed_alpha_rho_alpha} (iii), $\Fe_{\Phi}$
is a subalgebra in $T^*(\Me_{T^*})$ and $\rho^{it}\Fe_{\Phi}\rho^{-it}\subseteq \Fe_{\Phi}$
 for all $t\in \mathbb R$, (iii) implies (iv).

Suppose (iv) and let $\Ae_1$ be the subalgebra generated by $\{\rho^{it} d(\sigma,\rho)\rho^{-it},\ t\in \mathbb R,\ \sigma\in \Se\}$.
Then $\Ae_1\subseteq T^*(\Me_{T^*})$.
Let $E:\Ae\to \Ae_1$ be the trace preserving conditional expectation. Then its
adjoint is the embedding $E^*:\Ae_1\hookrightarrow \Ae$ and
since  $\rho^{it}\Ae_1\rho^{-it}\subseteq \Ae_1$ for all $t\in \mathbb R$, the map $E_\rho$ is the
$\rho$-preserving conditional expectation,
\cite{ac}. Hence
\[
E^*(d(E(\sigma),E(\rho)))=d(E(\sigma),E(\rho))=E_\rho(d(\sigma,\rho))=
d(\sigma,\rho).
\]
By the equivalence  of (ii) and (i) and Theorem \ref{thm:diverg} (viii)
 (for the map $E$),
 $E_\rho^*\circ E(\sigma)=\sigma$ for all $\sigma\in \Se$.

Let $F^*$  denote the embedding $\Me_{T^*}\hookrightarrow \Be$, then,  as above,
its adjoint $F=F^{**}:\Be\to \Me_{T^*}$ is the trace preserving conditional
expectation.
Let us define the map  $\bar T:\Me_{T^*}\to T^*(\Me_{T^*})$ by
$\bar T:= T^*\circ F^*$. Then since $T^*$ is faithful by Lemma
 \ref{lemma:faith}, $\bar T$ is injective,
so that $\bar T$ is a *-isomorphism and there is an inverse map
$R=(\bar T)^{-1}:T^*(\Me_{T^*})\to \Me_{T^*}$. Define the map
 $\hat S:\Be\to \Ae$  by $\hat S:= E^*_\rho\circ R^*\circ F$. Then $\hat S$ is completely positive and trace preserving. Moreover,
$T^*\circ \hat S^*=T^*\circ F^*\circ R \circ E_\rho=\bar T\circ R\circ E_\rho= E^*\circ E_\rho$, so that
$\hat S\circ T(\sigma)= (E^*\circ E_\rho)^*(\sigma)=\sigma$ and (v) holds.

Suppose (v). Let $\Se_0:=T(\Se)$ and let $\sigma_0=T(\sigma)$ for $\sigma\in \Se$. Then since $\hat S(\sigma_0)=\sigma$ and
$T\circ \hat S(\sigma_0)=T(\sigma)=\sigma_0$, the map
$\hat S$ is reversible for $\Se_0$.
Hence by Theorem \ref{thm:diverg} (viii),
the map $\hat T:=\hat S_{\rho_0}^*$ is completely positive and
 satisfies $\hat T(\sigma)=\sigma_0$, this proves (vi). The implication
 (vi) $\to $ (i) is clear.

\qed

\begin{rem} Note  that by the proof of (v),
the completely positive maps $\hat T$ and
 $\hat S$ can always be given as adjoints of a composition of a conditional
expectation and a *-isomorphism.

\end{rem}

\begin{coro}\label{coro:ext} Under the conditions 1 and 2, $T$ is reversible for
$\Se$
if and only if $T$ is reversible for $\tilde \Se:=\bigcup \{\rho^{is}\Se\rho^{-is},\
s\in \mathbb R\}$.

\end{coro}

{\it Proof.} Suppose $T$ is reversible for $\Se$. Let $\sigma\in \Se$ and
 let $d=d(\sigma,\rho)$. Then $d\in \Fe_{\Phi}$ and therefore also
$d(\rho^{is}\sigma\rho^{-is},\rho)=\rho^{is}d\rho^{-is}\in \Fe_{\Phi}$, for all $s\in \mathbb R$.

\qed

\subsection{Factorization}

In this Section, we give a characterization of reversibility in terms of the
structure of states in $\Se$. More precisely, we show that the elements in $\Se$
 must have the form of a product of two positive operators, such that $T^*$ is
 multiplicative on one of them and the other does not depend on $\sigma$. This can be viewed as a quantum version of the classical factorization theorem for sufficient statistics, see e.g. \cite{strasser}. The first such factorization result was proved in \cite{mope}, see also   \cite[Theorem 6.1]{bhmp}.
Similar conditions for the infinite dimensional case are proved in \cite[Theorem 6]{japetz}.

\begin{thm}\label{thm:factor} Assume  conditions 1 and 2.
Let $\Phi=T^*\circ T_\rho$ and $\tilde \Phi= T_\rho\circ T^*$.
Then  the  following are equivalent.
\begin{enumerate}
\item[(i)] $T$ is reversible for $\Se$.
\item[(ii)] There is a positive invertible element $\rho^B\in
\Fe_{\Phi}'\cap \Ae$, such that for each $\sigma \in \Se$,
\[
\sigma=T^*(\sigma_0^A)\rho^B,\quad T(\sigma)=\sigma_0^AT(\rho^B),
\]
with some  $\sigma_0^A\in \Fe_{\tilde \Phi}^+$.
\item[(iii)] There  is an element $\rho^B\in \Ae^+$, such that for each $\sigma \in \Se$,
\[
\sigma=T^*(\sigma_0^A)\rho^B,\quad T(\sigma)=\sigma_0^AT(\rho^B),
\]
with some  $\sigma_0^A\in \Be^+$.
\item[(iv)] There  is an element $\rho^B\in \Ae^+$, such that  each
$\sigma \in \Se$ has the form
\[
\sigma=\sigma^A\rho^B,
\]
where  $\sigma^A$ is a positive element in $T^*(\Me_{T^*})$.

\end{enumerate}

\end{thm}

{\it Proof.} Let us denote $\sigma_0:=T(\sigma)$ for $\sigma\in \Se$.
Suppose (i)  and let
\[
\rho=T^*(\rho_0^A)\rho^B,\qquad \rho_0=\rho_0^AT(\rho^B)
\]
be the decomposition from Theorem \ref{thm:fixed_alpha_rho_alpha} (vi).
 Then by Theorems \ref{thm:rev_d} and \ref{thm:fixed_alpha_rho_alpha}
 we have for $\sigma\in \Se$,
\begin{eqnarray*}
\sigma&=&\rho^{1/2} d(\sigma,\rho)\rho^{1/2}=\rho^{1/2}T^*(d(\sigma_0,\rho_0))
\rho^{1/2}
=T^*(\rho_0^A)^{1/2}T^*(d(\sigma_0,\rho_0))T^*(\rho_0^A)^{1/2} \rho^B\\
&=& T^*((\rho_0^A)^{1/2}d(\sigma_0,\rho_0)(\rho_0^A)^{1/2})\rho^B=
T^*(\sigma_0^A)\rho^B
\end{eqnarray*}
where we put  $\sigma^A_0:=(\rho_0^A)^{1/2}d(\sigma_0,\rho_0)(\rho_0^A)^{1/2}$.
Since $d(\sigma_0,\rho_0)=T_\rho(d(\sigma,\rho))\in
\Fe_{\tilde\Phi}^+$, $\sigma^A_0$ is a positive element in
$\Fe_{\tilde\Phi}$. Moreover,
$\sigma_0^A=T(\rho^B)^{-1/2}\sigma_0T(\rho^B)^{-1/2}$, hence
\[
\sigma_0=\sigma_0^AT(\rho^B),
\]
where we used Theorem \ref{thm:fixed_alpha_rho_alpha} (v). This proves (ii).
It is clear that (ii) implies (iii).

Suppose (iii). Then for $a\in \Be$,
\[
\Tr a\sigma_0  = \Tr a\sigma_0^A T(\rho^B)=\Tr T^*(a\sigma_0^A)\rho^B
\]
On the other hand,
\[
\Tr a \sigma_0=\Tr T^*(a)\sigma =\Tr T^*(a)T^*(\sigma_0^A)\rho^B
\]
Putting $a=\sigma_0^A$, we obtain
\[
\Tr T^*((\sigma_0^A )^2)\rho^B=\Tr T^*(\sigma_0^A)^2\rho^B
\]
Since $\rho$ is invertible, the decomposition implies that $\rho^B$ must be
invertible as well, hence by Schwarz inequality,
$T^*((\sigma_0^A )^2)=T^*(\sigma_0^A)^2$. This implies that
$\sigma_0^A\in \Me_{T^*}$, which proves (iv) with $\sigma^A:=T^*(\sigma_0^A)$.

Finally, suppose (iv). Let $\sigma\in \Se$.
Since both $\sigma^A$ and $\rho^B$ are positive
 and so is their product $\sigma$, they must commute. It follows that
\[
w_t:=\sigma^{it}\rho^{-it}=(\sigma^A)^{it}(\rho^A)^{-it}\in T^*(\Me_{T^*})
\]
for all $t\in \mathbb R$, where $\rho=\rho^A\rho^B$ is the decomposition for $\rho$. We have  $\rho^{is} w_t\rho^{-is}=w_s^*w_{t+s}\in T^*(\Me_{T^*})$
 for all $t,s\in \mathbb R$.  By analytic continuation for $t=-i/2$, we get
$\rho^{is}\sigma^{1/2}\rho^{-1/2}\rho^{-is}\in T^*(\Me_{T^*})$, hence also
$\rho^{is}d(\sigma,\rho)\rho^{-is}\in T^*(\Me_{T^*})$ for all $s$. By Theorem
\ref{thm:rev_d} (v), this implies (i).

\qed

The next Corollary shows that the recovery map $T_\rho$ does not depend on the
 choice of $\rho$. For faithful states, this was proved already in \cite{petz88}.

\begin{coro} Suppose the conditions 1 and 2 hold. Then  $T$ is reversible for
$\Se$ if and only if $T_\sigma= T_\rho|_{\supp\sigma \Ae \supp\sigma} $ for all $\sigma\in \Se$.
\end{coro}

{\it Proof.} Let $\sigma\in \Se$, $q:=\supp \sigma$, $q_0:=\supp T(\sigma)$
 and suppose that $T$ is
reversible for $\Se$. Let us denote $w=\sigma^{1/2}\rho^{-1/2}$,
$w_0=T(\sigma)^{1/2}T(\rho)^{-1/2}$.
By Theorem \ref{thm:factor} (ii) and Theorem \ref{thm:fixed_alpha_rho_alpha}, we have
\[
w_0=(\sigma_0^A)^{1/2}(\rho_0^A)^{-1/2}\in \Fe_{\tilde \Phi}
\]
and
\[
w=T^*(w_0)\in \Fe_\Phi,\qquad w_0=T_\rho(w)
\]
Then for $a\in q\Ae q$,
\begin{eqnarray*}
T_\sigma(a)&=& T(\sigma)^{-1/2}T(\sigma^{1/2}a\sigma^{1/2})T(\sigma)^{-1/2}\\
&=&
(w_0^{-1})^*T_\rho(w^*aw)w_0^{-1}=(w_0^{-1})^*T_\rho(w)^*T_\rho(a)T_\rho(w)w_0^{-1}\\
&=& q_0T_\rho(a)q_0
\end{eqnarray*}
Since $\rho^B$ is invertible, we must have $q_0=\supp \sigma^A_0\in
\Fe_{\tilde \Phi}$ and $q=\supp T^*(\sigma^A_0)=T^*(q_0)$. Hence also $T_\rho(q)=
q_0$ and
$q_0T_\rho(a)q_0=T_\rho(qaq)=T_\rho(a)$.

Conversely, since $T_\rho$ is unital, the equality
$T_\sigma= T_\rho|_{q \Ae q} $ implies that
$T_\sigma^*=T_\rho^*|_{q_0\Be q_0}$ by Lemma \ref{lemma:supp}, so that
$T_\rho^*\circ T(\sigma)=T_\sigma^*\circ T(\sigma)=\sigma$ and
 $T$ is reversible for $\Se$.

\qed

\subsection{Quantum hypothesis testing}

Let $\sigma$ and $\rho$ be density operators in $\Ae$. Let us consider
 the problem of testing the hypothesis $H_0=\rho$ against the alternative
$H_1=\sigma$. Any test  is represented by
an operator $0\le M\le I$, which corresponds to rejecting the hypothesis.
Then we have the error probabilities
\[
\alpha(M)=\Tr \rho M, \qquad \beta(M)=\Tr \sigma(1-M)
\]
For  $s\in [0,1)$, we define  the Bayes optimal test to be a minimizer
of the expression
\begin{equation}\label{eq:bayes}
s\alpha(M)+(1-s)\beta(M)= (1-s)(1-\Tr (\sigma-t\rho)M),\qquad t=\frac s{1-s}
\end{equation}
Then the minimal Bayes error probability  is
\[
\Pi_s:=\min_{0\le M\le I}\{ s\alpha(M)+(1-s)\beta(M)\} =
s\alpha(M_{\frac s{1-s}})+(1-s)\beta(M_{\frac s{1-s}})
\]
where $M_t$ maximizes the expression $\Tr (\sigma-t\rho)M$ over all $0\le M\le I$.
 Below we formulate the  quantum version of the Neyman-Pearson lemma. The obtained Bayes optimal tests are called the (quantum) NP tests for $(\rho,\sigma)$.
 
 If $a\in \Ae$ is a self adjoint operator, we denote by $a_+$ the positive part of $a$, that is, $a_+=\sum_{i,\lambda_i>0}p_i$, where $a=\sum_i\lambda_ip_i$ is the spectral decomposition of $a$.
 
\begin{lemma}\label{lemma:qnp}\cite{holevo,helstrom}
For  $t\ge 0$, let $P_{t,+}:=\supp (\sigma-t\rho)_+$ and let $P_{t,0}$ be the
projection onto the kernel of $\sigma-t\rho$. Then
$0\le M_t\le I$ is a Bayes optimal test if and only if
\[
M_t=P_{t,+}+X_t
\]
with $0\le X_t\le P_{t,0}$.
The minimal Bayes error probability  is
\[
\Pi_s = \frac 12(1-\|(1-s)\sigma- s\rho\|_1)
\]
\end{lemma}

Let now $T: \Ae\to \Be$ be a trace preserving positive map.
 Let $s\in(0,1)$, $t=s(1-s)^{-1}$ and let
$\Pi^0_s$ be the minimal  Bayes error probability for testing the hypothesis
$H_0=T(\rho)$ against $H_1=T(\sigma)$. For $N\in \Be$,
 $0\le N\le I$, we have
\[
\Tr (T(\sigma)-tT(\rho))N=\Tr (\sigma-t\rho)T^*(N)\le \max_{0\le
M\le I}\Tr  (\sigma-t\rho)M
\]
so that $\Pi_s^0\ge \Pi_s$, this is equivalent to the fact that
\begin{equation}\label{eq:norm_1monot}
\| T(\sigma-t\rho)\|_1\le \|\sigma-t\rho\|_1
\end{equation}

In \cite{ja2suf}, equality in (\ref{eq:norm_1monot}) was
investigated for a pair of invertible density operators, in the case when $T$ is
 the restriction to a subalgebra. If equality holds for all $t\ge 0$, then the subalgebra must contain some Bayes optimal test for all $s\in [0,1]$, such subalgebras are called
2-sufficient. It was shown that in some cases, 2-sufficiency is equivalent to sufficiency, that is, reversibility of $T$ for $\{\sigma,\rho\}$.
From another point of view, this condition was  studied also in \cite{kohout}
and it was shown that for a completely positive trace preserving map, 
the equality implies reversibility for certain sets $\Se$. 

Since the $L_1$-norm is one of the basic distance measures on states, equivalence between equality in (\ref{eq:norm_1monot}) and reversibility is an important open question. We will show below (Theorem \ref{thm:suff2_d}) that this 
equivalence holds if equality in (\ref{eq:norm_1monot}) is required for all
$\sigma$ in the extended family $\tilde \Se= \bigcup
\{\rho^{is}\Se\rho^{-is},\ s\in \mathbb R\}$. Moreover, Theorem \ref{thm:qcbhb} shows this
 equivalence if equality in (\ref{eq:norm_1monot}) holds for $n$  copies of the states, for all $n$.

We will suppose below that $\rho$ is invertible.

\begin{lemma}\label{lemma:eigen}\cite[Lemma 4]{ja2suf}.
$P_{t,0}\ne 0$ if and only if $t$
is an eigenvalue of $d(\sigma,\rho)$.
Moreover, the rank of $P_{t,0}$ is equal to the multiplicity of $t$.

\end{lemma}

\begin{lemma}\label{lemma:ctns}
 The function $t\mapsto P_{t,+}$ is right-continuous.
Moreover,
\[
\lim_{s\to t^-} P_{s,+}=P_{t,+}+P_{t,0},\qquad t\ge 0.
\]

\end{lemma}

{\it Proof.}
Let $\rho(t):=\sigma-t\rho$ for $t\in \mathbb R$. Let
$\lambda^\downarrow_1(t),\dots,\lambda^\downarrow_N(t)$ denote the decreasingly
ordered eigenvalues of $\rho(t)$ (with multiplicities). For
$t_1,t_2\in \mathbb R$, we have $\rho(t_1)=\rho(t_2)+(t_2-t_1)\rho$.
By Weyl's perturbation theorem \cite[Corollary III.2.6]{bhatia_ma}, this implies that
\[
\max_j|\lambda^\downarrow_j(t_1)-\lambda^\downarrow_j(t_2)|\le |t_1-t_2|\|\rho\|.
\]
 Moreover,
since $\rho$ is invertible, we obtain  by \cite[Corollary III.2.2]{bhatia_ma}
  that
\[
\lambda^\downarrow_j(t_2)<\lambda^\downarrow_j(t_2)+(t_2-t_1)\lambda^\downarrow_N(\rho)\le \lambda^\downarrow_j(t_1)
\]
when $t_1<t_2$, where $\lambda^\downarrow_N(\rho)$ denotes the smallest eigenvalue of $\rho$. Hence the functions $t\mapsto \lambda^\downarrow_j(t)$
are continuous and strictly decreasing.

It is clear that for $t<0$ all $\lambda^\downarrow_j(t)$ are strictly positive, and that $\lambda^\downarrow_j(t)=0$ for some index $j$ if and only if
 $P_{t,0}\ne 0$. Let $0\le t_1<\dots<t_n$ be the eigenvalues of $d(\sigma,\rho)$ and put $t_0:=0$, $t_{n+1}:=\infty$. Then there are indices $i_k\in\{1,\dots,N\}$, $k=1,\dots,n$, such that $N=i_1>i_2>\dots>i_n>i_{n+1}:=0 $ and for every $t\in [t_{k-1},t_k)$
 the strictly positive eigenvalues of $\rho(t)$ are given by
$\lambda^\downarrow_1(t),\dots,\lambda^\downarrow_{i_k}(t)$.

Let $t\in [t_{k-1},t_k)$ and let $\gamma(t)$ be a circle, contained entirely  in the open half-plane of complex numbers having
strictly positive real parts and enclosing all $\lambda^\downarrow_1(t),\dots,\lambda^\downarrow_{i_k}(t)$.
 By continuity of  $\lambda^\downarrow_j$, there is some $\delta>0$ such that $\gamma(t)$ encloses
 $\lambda^\downarrow_1(s),\dots,\lambda^\downarrow_{i_k}(s)$ for all $s\in (t-\delta,t+\delta)$ and
 $[t,t+\delta)\subset[t_{k-1},t_{k})$. Then
\[
P_{s,+}=\frac 1{2i\pi}\oint_{\gamma(t)}(zI-\rho(s))^{-1}dz,\qquad s\in
[t, t+\delta).
\]
This implies that $t\mapsto P_{t,+}$ is right-continuous.
Let now $t\in (t_{k-1},t_k)$, then we can find $\delta>0$ as above, but such that, moreover, $(t-\delta,t+\delta)\subset
(t_{k-1},t_k)$. In this case,
\[
P_{s,+}=\frac 1{2i\pi}\oint_{\gamma(t)}(zI-\rho(s))^{-1}dz,\qquad s\in (t-\delta, t+\delta)
\]
so that $t\mapsto P_{t,+}$ is continuous at $t$. Suppose $t=t_{k-1}$, then by definition of $i_k$ and $t_{k-1}$, we must have
\[
\lambda^\downarrow_j(t_{k-1})\left\{\begin{array}{cc} > 0 & j\le i_k\\
=0 & j=i_k+1,\dots,i_{k-1}\\
<0 & j>i_{k-1}
\end{array}\right.
\]
Let $\gamma'_k$ be a circle in the complex plane, enclosing $\lambda^\downarrow_1(t_{k-1}),\dots,\lambda^\downarrow_{i_k}(t_{k-1})$
and 0, but such that the closed disc encircled by $\gamma'_k$ does not contain any other eigenvalue of $\rho(t_{k-1})$.
Then there is some
$\delta>0$  such that $(t_{k-1}-\delta,t_{k-1})\subset [t_{k-2},t_{k-1})$ and
\[
P_{s,+}=\frac 1{2i\pi}\oint_{\gamma'_{k}}(zI-\rho(s))^{-1}dz,\qquad s\in
(t_{k-1}-\delta,t_{k-1})
\]
 It follows that $\lim_{s\to t_{k-1}^-}P_{s,+}=P_{t_{k-1},+}+P_{t_{k-1},0}$. Since
 $P_{t,0}=0$ for $t\notin\{t_1,\dots,t_n\}$, this proves the assertion.

\qed

Let us denote $Q_{t,+}:=\supp (T(\sigma)-tT(\rho))_+$ and
$Q_{t,0}$ the projection onto the kernel of $T(\sigma)-tT(\rho) $.

\begin{lemma}\label{lemma:norm_1} Let $T: \Ae\to \Be$ be a trace preserving
 positive map and suppose that
both $\rho$ and $T(\rho)$ are invertible.
 The following are equivalent.
\begin{enumerate}
\item [(i)] $\| T(\sigma)-tT(\rho)\|_1= \|\sigma-t\rho\|_1$, for all $t\in \mathbb R$.
\item[(ii)] $P_{t,+}=T^*(Q_{t,+})$, $P_{t,0}=T^*(Q_{t,0})$ for $t\in \mathbb R$.
\end{enumerate}

\end{lemma}

{\it Proof.} Since  $Q_{t,+}$ is  an NP test for $(T(\rho),T(\sigma))$,  (i) implies that
\[
\Tr (T(\sigma)-tT(\rho))Q_{t,+}=\Tr (\sigma-t\rho)T^*(Q_{t,+})=\max_{0\le M\le I} \Tr (\sigma-t\rho)M
\]
so that $T^*(Q_{t,+})$ is an NP test for $(\rho,\sigma)$. By Lemma \ref{lemma:qnp}, there is  some
$0\le X_t\le P_{t,0}$, such that
$T^*(Q_{t,+})=P_{t,+}+X_t$.
 It follows that $P_{t,+}=T^*(Q_{t,+})$ holds for all $t$ such that $P_{t,0}=0$, that is, for $t\in \mathbb R\setminus\{t_1,\dots,t_n\}$.
Since $t\mapsto P_{t,+}$ and $t\mapsto T^*(Q_{t,+})$ are right continuous, it follows that $T^*(Q_{t,+})=P_{t,+}$
for all $t$. On the other hand, by Lemma \ref{lemma:ctns} we have for all $t$
\[
P_{t,+}+P_{t,0}=\lim_{s\to t^-} P_{s,+}=\lim_{s\to t^-}T^*(Q_{s,+})=T^*(Q_{t,+})+
T^*(Q_{t,0})
\]
hence $P_{t,0}=T^*(Q_{t,0})$ for all $t$.
The converse is obvious.

\qed

\begin{thm}\label{thm:suff2_d} Assume  the conditions 1 and 2. Then
\begin{enumerate}
\item[(i)] $T$ is reversible for $\Se$ if and only if
\begin{equation}\label{eq:norm_1modular}
\|\sigma-t\rho\|_1=\|T(\sigma)-tT(\rho)\|_1,\qquad \sigma\in\tilde \Se,\ t\ge 0
\end{equation}
\item[(ii)] Suppose that $\rho^{is}\Se\rho^{-is}\subseteq \Se$ for all $s\in \mathbb R$. Then $T$ is reversible
for $\Se$ if and only if
\begin{equation}\label{eq:norm1}
\|\sigma-t\rho\|_1=\|T(\sigma)-tT(\rho)\|_1,\qquad \sigma\in \Se,\  t\ge 0
\end{equation}
\item[(iii)] Suppose that $\Be$ is abelian. Then $T$ is reversible for $\Se$ if and only if
(\ref{eq:norm1}) holds. Moreover, in this case all elements in $\Se$ commute.
\item [(iv)] Suppose that all elements in $\Se$ commute with $\rho$. Then
$T$ is reversible for $\Se$ if and only if
(\ref{eq:norm1}) holds.
\end{enumerate}
\end{thm}

{\it Proof.} (i) By Corollary \ref{coro:ext}, $T$ is reversible for $\Se$ if and only if it is reversible for   $\tilde \Se$.
By monotonicity (\ref{eq:norm_1monot}), we get (\ref{eq:norm_1modular}).

For the converse, let $\sigma\in \tilde \Se$. Then by Lemma \ref{lemma:norm_1},
 (\ref{eq:norm_1modular}) implies that $P_{t,0}=T^*(Q_{t,0})$ for the corresponding projections for $\sigma$ and $\rho$.  This implies that
$Q_{t,0}\in \Me_{T^*}$ and $P_{t,0}\in T^*(\Me_{T^*})$.

Let $t_1,\dots,t_n$ be the eigenvalues of $d=d(\sigma,\rho)$ and let
$F_1,\dots,F_n$ be the corresponding eigenprojections. Denote $P_i:=P_{t_i,0}$.
Then we have
$(d-t_i)\rho^{1/2}P_i=\rho^{-1/2}(\sigma-t_i\rho)P_i=0$
and this implies
\[
d\rho^{1/2}\sum_iP_i=\rho^{1/2}\sum_it_iP_i
\]
Moreover, any vector in the range of $\rho^{1/2}P_i\rho^{1/2}$ is an eigenvector
of $d$, so that $\supp(\rho^{1/2}P_i\rho^{1/2})\le F_i$ and  by Lemma
\ref{lemma:eigen}, $\mathrm{rank}(F_i)=\mathrm{rank}(P_i)=\mathrm{rank}(\rho^{1/2}P_i\rho^{1/2})$. It follows that
$\sum_i P_i$ is invertible, so that
$d(\sigma,\rho)=\rho^{1/2}c\rho^{-1/2}$, with
\[c:=\sum_i t_i P_i(\sum_j P_j)^{-1}\in T^*(\Me_{T^*}).\]
It follows that for $s\in \mathbb R$ and $\sigma\in \Se$,
$\rho^{is-1/2}d(\sigma,\rho)\rho^{1/2-is}\in T^*(\Me_{T^*})$.
By analytic continuation, we get $\rho^{it}d(\sigma,\rho)\rho^{-it}\in
T^*(\Me_{T^*})$ for all $t\in \mathbb R$, which implies that $T$ is reversible for $\Se$, by Theorem
\ref{thm:rev_d}.

(ii) clearly follows from (i).

(iii) Let $\sigma\in \Se$ and let $P_{t,0}$ and $Q_{t,0}$ be the corresponding projections.
Note that since $\Be$ is commutative, $Q_{t,0}$ must commute for all $t$.
Suppose that (\ref{eq:norm1}) holds, then $P_{t,0}=T^*(Q_{t,0})$ and, since then
$Q_{t,0}\in \Me_{T^*}$,  this implies that all
$P_{t,0}$ commute as well.
As in the proof of (i), $d(\sigma,\rho)=\rho^{1/2}c\rho^{-1/2}$,
where we now have $c\ge0$. This implies that $d(\sigma,\rho)\rho=\rho^{1/2} c\rho^{1/2}\ge0$,
hence   $d(\sigma,\rho)\rho=\rho d(\sigma,\rho)$ and therefore also $\sigma\rho=\rho\sigma$.
This implies that $\rho^{is}\sigma\rho^{-is}=\sigma$ and the statement follows by (ii).
 The converse implication is clear.

(iv) follows from (ii).

\qed

\subsection{Quantum Chernoff and Hoeffding distances}

Let $n\in \mathbb N$ and suppose we are given $n$ identical copies of the states
$\rho^{\otimes n},\sigma^{\otimes n}\in \Se(\Ae^{\otimes n})$.
Consider the problem of testing the hypothesis $H_0=\rho^{\otimes n}$ against
 $H_1=\sigma^{\otimes n}$. Then the minimum Bayes error probability is
\[
\Pi_{s,n}=\frac12(1-\|(1-s)\sigma^{\otimes n}-s\rho^{\otimes n}\|_1)
\]
It is an important result of \cite{auden} that as $n\to \infty$, the probabilities $\Pi_{s,n}$ decay exponentially fast and the rate of convergence is given by
\begin{equation}\label{eq:qcb}
\lim_n-\frac 1n\log \Pi_{s,n}=-\log\left(\inf_{0\le u\le 1}\Tr\sigma^u\rho^{1-u}
\right)=:C(\sigma,\rho)
\end{equation}
for any $s\in [0,1]$, where we put $x^0=\supp x$ for any positive
$x\in \Ae$.   The quantity $C(\sigma,\rho)$ is called the
 quantum Chernoff distance. Note that  $C$ is related to the
 convex quantum $f$-divergence $S_u(\sigma,\rho)$, but one can show that
$C$ itself is not an $f$-divergence \cite{bhmp}. Nevertheless,
if $T$ is the adjoint of a unital Schwarz map, then $C$ satisfies monotonicity:
\[
C(\sigma,\rho)\ge C(T(\sigma),T(\rho))
\]
and, moreover,  $C(\sigma,\rho)=0$ if and only if $\sigma=\rho$.

Let us consider again the problem of testing the hypothesis $H_0=\rho$ against the alternative $H_1=\sigma$. Let $0\le M\le I$ be a test.
Differently from the Bayesian approach, in the asymmetric approach the error
probability $\alpha(M)$ is bounded, $\alpha(M)\le \epsilon$ for some fixed
$\epsilon >0$. The error probability $\beta(M)$ is then minimalized over all
tests, under this constraint,
\[
\beta_\epsilon:=\inf\{ \beta(M),\ 0\le M\le I,\ \alpha(M)\le \epsilon\}.
\]
Suppose we have $n$ independent copies of the states $\sigma^{\otimes n}$ and
$\rho^{\otimes n}$ and let $M_n\in\Ae^{\otimes n} $.
Here we require that the probabilities $\alpha(M_n)$ decay exponentially as $n\to \infty$. Let $r> 0$ and put
\[
\beta_{r,n}:=\inf\{ \beta(M_n),\ 0\le M_n\le I,\ \alpha(M_n)\le e^{-nr}\}
\]
The following equality  was proved in \cite{hayashi} and \cite{nagaoka}:
For $r>0$,
\[
\lim_n -\frac 1n \log \beta_{r,n}=\sup_{0\le u<1} \frac{-ur-\log \Tr \rho^u\sigma^{1-u}}{1-u}=:H_r(\rho,\sigma)
\]
The limit expression is called  the quantum Hoeffding distance.
Similarly as the Chernoff distance, $H_r$ is not an $f$-divergence \cite{bhmp},
but it is related to $S_u$.  This implies the monotonicity
\[
H_r(T(\sigma),T(\rho))\le H_r(\sigma,\rho)
\]
for  $T$ the adjoint of a unital Schwarz map.
Moreover, by \cite{hmo}, see also \cite{bhmp},
\[
H_0(\sigma,\rho):=\lim_{r\to 0}H_r(\sigma,\rho)= S(\sigma,\rho)=
\Tr \sigma(\log\sigma-\log\rho)
\]
holds if $\supp\sigma\le\supp\rho$.

Suppose that $q:=\supp\sigma\le\supp\rho$, then the function
$[0,\infty)\ni r\mapsto
H_r(\sigma,\rho)$ has the following properties \cite{hmo}, see also
\cite{auden}:

The function  is convex and lower semicontinuous,
 for $r\in [0, S_\sigma(\rho,\sigma)]$ it is strictly convex and decreasing,
and for $r\ge S_\sigma(\rho,\sigma)$ it has a constant value $H_r(\sigma,\rho)=
-\log \Tr q\rho$, here
\[
S_\sigma(\rho,\sigma)=-\log\Tr q\rho+\frac 1{\Tr q\rho}\Tr \rho(\log\rho-\log\sigma)q
\]
Note that if $\supp \sigma= \supp \rho$, then $S_\sigma(\rho,\sigma)=S(\rho,\sigma)$.

\begin{prop}\label{prop:qcbhb}\cite{bhmp} Let $\sigma$ and $\rho$ be two density
operators in $\Ae$ such that $\supp \sigma=\supp \rho$. Let $T: \Ae\to \Be$ be
the adjoint of a unital Schwarz map and suppose that one of the following conditions holds:
\begin{enumerate}
\item[(i)]  $C(\sigma,\rho)=C(T(\sigma),T(\rho))$
\item[(ii)] $H_r(\sigma,\rho)=H_r(T(\sigma),T(\rho))$ for some $r\in [0,S(T(\rho),T(\sigma))]$
\end{enumerate}
Then $T_\rho^*\circ T(\sigma)=\sigma$.

\end{prop}

\begin{thm}\label{thm:qcbhb} Assume the conditions 1 and 2. Then the following
are equivalent.
\begin{enumerate}
\item[(i)] $T$ is reversible for $\Se$.
\item[(ii)]  $C(\sigma,\rho)=C(T(\sigma),T(\rho))$
 for all $\sigma\in co(\Se)$.
\item[(iii)] $\|\sigma^{\otimes n}-t\rho^{\otimes n}\|_1=\|T(\sigma)^{\otimes n}-
tT(\rho)^{\otimes n}\|_1$ for all $\sigma\in \Se$, $t\ge 0$ and $n\in \mathbb N$.
\item[(iv)] $H_r(\sigma,\rho)=H_r(T(\sigma),T(\rho))$ for all $\sigma\in \Se$ and $r\ge 0$.
\end{enumerate}
Suppose moreover that there is some  $\Se_0\subset \Se$, such that
$\Se\subseteq \overline{co}(\Se_0\cup\{\rho\})$ and
$T(\rho)\notin \overline{T(\Se_0)}$.  Then there exists some $r_0>0$ such that
(i) - (iv) are equivalent to
\item[(v)] $H_r(\sigma,\rho)=H_r(T(\sigma),T(\rho))$ for
all $\sigma\in co(\Se)$ and  some $r\in [0,r_0]$.

\end{thm}

{\it Proof.} Since $T$ is reversible for $\Se$ if and only if it is reversible for $co(\Se)$, (i) implies (ii) by monotonicity of $C$. Conversely, suppose (ii) and let $\sigma\in \Se$, then $\sigma_1:=\tfrac 12(\sigma+\rho)$ is an invertible element in $co(\Se)$. Proposition \ref{prop:qcbhb} now implies that
  $T^*_\rho \circ T(\sigma_1)=\sigma_1$ and by (\ref{eq:alpha_rho_preserves}),
 we have also $T^*_\rho\circ T(\sigma)=\sigma$.

Further, suppose (i),
then by Theorem \ref{thm:rev_d} (vi),
there are trace preserving  completely positive maps $\hat T:\Ae\to \Be$
 and $\hat S:\Be\to \Ae$, such that $\hat T(\sigma)=T(\sigma)$,
$\hat S\circ T(\sigma)=\sigma$, $\sigma\in \Se$. It follows that
$T(\sigma)^{\otimes n}=\hat T(\sigma)^{\otimes n}=\hat T^{\otimes n}(\sigma^{\otimes n})$ and $\sigma^{\otimes n}=\hat S^{\otimes n}(T(\sigma)^{\otimes n})$,
 for all $\sigma\in \Se$,
  where
$\hat T^{\otimes n}$ and $\hat S^{\otimes n}$ are
 completely positive and trace preserving. By monotonicity of the $L_1$-norm,
 this implies (iii). The implications (iii) $\implies $ (iv) $\implies$ (i) were
already proved in \cite{bhmp}.

Suppose now that the additional condition holds.
Let us  choose some $\varepsilon\in (0,1)$ and
put
\[
r_0:=\inf_{\sigma\in \Se_0} S(T(\rho),T(\varepsilon\rho+(1-\varepsilon)\sigma)).
\]
 Then if
$r_0=0$, there exists a
 sequence $\sigma_n\in \Se_0$, such that
$S(T(\rho),T(\varepsilon\rho+(1-\varepsilon)\sigma_n))\to 0$. This implies that
 $T(\sigma_n)\to T(\rho)$, so that  $T(\rho)\in \overline{ T(\Se_0)}$,
which is not possible.
 Hence $r_0>0$.

Suppose (v) holds and let $\sigma\in \Se_0$. Denote
$\sigma_\varepsilon=\varepsilon\rho+(1-\varepsilon)\sigma$. Then
$0\le r\le S(T(\rho),T(\sigma_\varepsilon))$.  Since
$\sigma_\varepsilon$ is invertible, we can apply Proposition  \ref{prop:qcbhb},
which implies that   $T_\rho^*\circ T(\sigma_\varepsilon)=\sigma_\varepsilon$
and therefore also
$T_\rho^*\circ T(\sigma)=\sigma$ for all $\sigma\in \Se_0$.
Since $\Se\subseteq \overline{co}(\Se_0\cup\{\rho\})$, this implies (i).
The implication (i) $\implies$ (v) follows by monotonicity.

\qed

\begin{rem} Note that if all elements in $\Se$ are invertible, then we may
replace $co(\Se)$ by $\Se$ in (ii) and by $\Se_0$ in (v), where we put
$r_0:=\inf_{\sigma\in \Se_0} S(T(\rho),T(\sigma))$.
\end{rem}

\subsection{Quantum Fisher information and $\chi^2$-divergence}

Let us denote by $\mathcal D$  the set of invertible
density operators in $\Ae$. Then $\mathcal D$ is a differentiable manifold, where the tangent space
 at each point $\rho\in \mathcal D$ is the vector space $\mathcal T_\rho$ of traceless self-adjoint elements in $\Ae$.

A monotone  metric on $\mathcal D$ is a Riemannian metric $\lambda_\rho$, satisfying
\begin{equation}\label{eq:monot_metric}
\lambda_\rho(x,x)\ge \lambda_{T(\rho)}(T(x),T(x)),\qquad x\in \mathcal T_\rho,\ \rho\in \mathcal D
\end{equation}
for any completely positive trace preserving  map $T:\Ae\to \Be$. 

It was proved by Petz in \cite{petz96} that any
monotone metric has the form
\[
\lambda_\rho(x,y)=\Tr (J^f_\rho)^{-1}(x)y
\]
with $J_\rho^f= f(\Delta_\rho)R_\rho$, where $\Delta_\rho:=\Delta_{\rho,\rho}=
L_\rho R_\rho^{-1}$, and $f:(0,\infty)\to (0,\infty)$ an operator monotone function satisfying the symmetry $f(t)=tf(t^{-1})$.
Under the normalization condition $f(1)=1$, the restriction of $\lambda_\rho$ to the submanifold of diagonal elements in $\mathcal D$
 coincides with the classical Fisher information for probability measures on a finite set, moreover, the monotonicity condition (\ref{eq:monot_metric})
 characterizes the classical Fisher information up to multiplication by a 
constant. Accordingly, any monotone metric with the above normalization is called a quantum Fisher information.

The operator $J_\rho^f$ satisfies
\cite{petz86b,OP}
\[
J_{T(\rho)}^f\ge TJ_\rho^fT^*
\]
for any operator monotone (not necessarily symmetric or normalized) function
$f$ and
$T:\Ae\to \Be$ the adjoint of a unital Schwarz map. This is equivalent to
\cite{petz96}
\begin{equation}\label{eq:monot_map}
(J_\rho^f)^{-1}\ge T^*(J_{T(\rho)}^f)^{-1}T,
\end{equation}
which implies that the monotonicity (\ref{eq:monot_metric}) holds for all such $f$ and $T$.

A related quantity is the quantum version of the $\chi^2$-divergence, which
was introduced in  \cite{rusetal}  as
\[
\chi_{1/f}^2(\sigma,\rho)=\lambda_\rho^f(\sigma-\rho,\sigma-\rho)
\]
where $\lambda_\rho^f$ is a monotone metric.

 Let now $f:(0,\infty)\to (0,\infty)$ be operator monotone. Then $t\mapsto f(t)^{-1}$ is a non-negative
 operator monotone decreasing function on $(0,\infty)$.
By \cite{hansen}, for each such function there is a
positive Borel measure $\nu_f$ with support in $[0,\infty)$ and
$\int_0^\infty (1+s^2)^{-1}d\nu_f(s)<\infty$,
$\int_0^\infty s(1+s^2)^{-1}d\nu_f(s)<\infty$, such that
\[
f(t)^{-1}=\int_0^\infty\frac 1{s+t}d\nu_f(s)=\int_0^\infty f_s(t)^{-1}d\nu_f(s)
\]
where  $f_s(t)=s+t$, $t\in \mathbb R^+$. Then it follows that
\begin{equation}\label{eq:monot_integral}
(J^f_\rho)^{-1}= f(L_\rho R_\rho^{-1})^{-1}R_\rho^{-1}=\int_0^\infty (s R_\rho+L_\rho)^{-1}d\nu_f(s)=\int_0^\infty
(J_\rho^s)^{-1}d\nu_f(s)
\end{equation}
where $J_\rho^s:=J^{f_s}_\rho=sR_\rho+L_\rho$.

\begin{lemma}\label{lemma:fisher} Let
$T:\Ae\to \Be$ be  the adjoint of a unital Schwarz map. Let $x\in \Ae$.
Then for $s\ge0$,
\begin{equation}\label{eq:monot_lam}
\Tr x^* (s R_\rho+L_\rho)^{-1}(x)\ge \Tr T(x)^*(s R_{T(\rho)}+L_{T(\rho)})^{-1}
(T(x))
\end{equation}
and equality holds if and only if
\begin{equation}\label{eq:equal_lam}
(s R_\rho+L_\rho)^{-1}(x)=T^*[(s R_{T(\rho)}+L_{T(\rho)})^{-1}(T(x))]
\end{equation}
\end{lemma}

{\it Proof.}
Since the function $f_s$ is operator monotone, the inequality
(\ref{eq:monot_lam}) follows from (\ref{eq:monot_map}) for $f=f_s$.
 If equality holds for some
$x\in \Ae$, then
\[
\<x,((J^s_\rho)^{-1}-T^*(J^s_{T(\rho)})^{-1}T)(x)\>=0
\]
which again by (\ref{eq:monot_map}) is equivalent to
$((J^s_\rho)^{-1}-T^*(J^s_{T(\rho)})^{-1}T)(x)=0$.

\qed

It follows from the above Lemma and the integral representation (\ref{eq:monot_integral}) that
$\lambda^f_\rho(x,x)=\lambda^f_{T(\rho)}(T(x),T(x))$ if and only if
(\ref{eq:equal_lam}) holds for all $s\in {\rm supp}\, \nu_f$, that is,
\begin{equation}\label{eq:equality_fish}
(s+\Delta_\rho)^{-1}(x\rho^{-1})=T^*[(s+\Delta_{T(\rho)})^{-1}(T(x)T(\rho)^{-1})],\quad s\in {\rm supp}\, \nu_f
\end{equation}

Let now $x\in \mathcal T_\rho$. Then since $\rho$ is
invertible, there exists some interval $I\ni 0$ such that
$\sigma_u:=\rho+ux\in \Se(\Ae)$ for $u\in I$. Let us denote by $I_{\rho,x}$
the largest such interval and let  $\Se_{\rho,x}:=\{\sigma_u, u\in I_{\rho,x}\}$.

\begin{prop}\label{prop:fisher}
Let $\rho\in \mathcal D$, $x\in\mathcal T_\rho$ and $T:\Ae\to \Be$  be such that
$T$ and $\Se_{\rho,x}$ satisfy the conditions 1 and 2.
Then the following are equivalent.
\begin{enumerate}
\item [(i)] $\lambda_\rho^f(x,x)=\lambda_{T(\rho)}^f(T(x),T(x))$ for a monotone
metric such that  $|{\rm supp}\, \nu_f |\ge |spec (\Delta_\rho)\cup spec(\Delta_{T(\rho)})|$
\item [(ii)] $\rho^{it}x\rho^{-it-1}=T^*(T(\rho)^{it}T(x)T(\rho)^{-it-1})$, $t\in \mathbb R$
\item [(iii)] $\rho^{-1/2}x\rho^{-1/2}=T^*(T(\rho)^{-1/2}T(x)T(\rho)^{-1/2})$
\item[(iv)] $T$ is reversible for $\Se_{\rho,x}$.
\item[(v)] $\lambda_\rho^f(x,x)=\lambda_{T(\rho)}^f(T(x),T(x))$ for any monotone metric $\lambda_\rho^f$.

\end{enumerate}
\end{prop}

{\it Proof.} Note that by the assumptions, $T(\rho)$ must be invertible. Suppose (i), then (\ref{eq:equality_fish}) holds and
by \cite[Lemma 5.2]{bhmp}, this implies that
\[
h(\Delta_\rho)x\rho^{-1}=h(\Delta_{T(\rho)})T(x)T(\rho)^{-1}
\]
for any complex-valued function $h$ on $spec(\Delta_\rho)\cup
spec(\Delta_{T(\rho)})$. In particular, for $h(\lambda)=\lambda^{it}$, we get (ii). We have  (ii) $\implies$ (iii) by analytic continuation for $t=i/2$.
Suppose (iii) and let
$\sigma_u\in \Se_{\rho,x}$. Then
\[
T^*(d(T(\sigma_u),T(\rho)))=T^*(I+uT(\rho)^{-1/2}T(x)T(\rho)^{-1/2}) =
I+ u\rho^{-1/2}x\rho^{-1/2}  =d(\sigma_u,\rho)
\]
 and by Theorem \ref{thm:rev_d}, this implies (iv).
(iv) implies (v) by monotonicity of Fisher information. The implication (v) $\implies$ (i) is trivial.

\qed

Let $\Se\subset\Se(\Ae)$ and let  $Lin(\Se)={\rm span}\,\{\sigma_1-\sigma_2:\ \sigma_1,\sigma_2\in \Se\}$. Then $Lin(\Se)$
 is a vector subspace in the real vector space of self-adjoint traceless
operators.

\begin{thm}\label{thm:fish} Suppose that the conditions 1 and 2 hold. Then the
following are equivalent.
\item[(i)] $T$ is reversible for $\Se$.
\item[(ii)] $\lambda_\rho^f(x,x)=\lambda_{T(\rho)}^f(T(x),T(x))$ for  all $x\in Lin (\Se)$ and  all monotone metrics.
\item[(iii)] $\chi^2_{1/f}(\sigma,\rho)=\chi^2_{1/f}(T(\sigma),T(\rho))$ for
all $\sigma\in \Se$ and  all $\chi^2$-divergences.
\item[(iv)] The equality in (ii) holds for
 some symmetric positive  operator monotone function $f$
 such that  $|{\rm supp}\, \mu_f|\ge \dim(\Ha)^2+\dim(\Ka)^2$.
\item[(v)] The equality in (iii) holds for some $f$ as in (iv).
\end{thm}

{\it Proof.} (i) implies (ii) by monotonicity of Fisher information and the
implication (ii) $\implies$ (iii) is clear. We also have (ii)
$\implies$ (iv) and both (iv) and (iii) imply (v). It is therefore enough to prove (v) $\implies$ (i). So suppose (v) and let $\sigma \in \Se$.
Put $x=\sigma-\rho$ in Proposition \ref{prop:fisher} (iii), then it follows that
$T^*(d(T(\sigma),T(\rho)))=d(\sigma,\rho)$ for $\sigma\in \Se$ which implies (i) by Theorem \ref{thm:rev_d}.

\qed

\begin{rem} An important example of a quantum Fisher information, 
resp. $\chi^2$-divergence, is given by $f(t)=\frac12 f_1(t)=\frac 12(1+t)$. 
In this case, $\nu_f$ is concentrated in $t=1$ and
$\lambda^f_\rho(x,y)=2\Tr y(L_\rho+R_\rho)^{-1}(x)$ is called the 
Bures metric. It is the smallest element in the family of quantum Fisher 
informations. The simple example below shows that preservation of the Bures metric
 does not imply reversibility, so that, once again, the support condition in (iv) resp. (v) of the above theorem cannot be dropped.

So let $y=y^*\in \Ae$ be such that $\rho y\ne y\rho$ and $\Tr \rho y=0$, and let 
$\mathcal C\subset \Ae$ be the  commutative subalgebra generated by $y$.
Then $z:=\rho y+y\rho\in \mathcal T_\rho$ and, by replacing $y$ by $ty$ for some $t>0$ if necessary, we may suppose that $\sigma:=\rho+z\in \mathcal D$.
Let $T:\Ae\to \mathcal C$ be the trace preserving conditional expectation, then 
$T(\sigma)=T(\rho)+T(z)=T(\rho)+T(\rho)y+yT(\rho)$. This implies that
\[
(L_\rho+R_\rho)^{-1}(\sigma-\rho)=y=(L_{T(\rho)}+R_{T(\rho)})^{-1}(T(\sigma)-T(\rho))
\]
which  implies that 
$\chi^2_{1/f}(\sigma,\rho) =\chi^2_{1/f}(T(\sigma),T(\rho))$. 

On the other hand, if $T$ is reversible, then by Theorem \ref{thm:factor} (iv),
$\rho$ and $\sigma$ must commute.
But we have   
$[\sigma,\rho]=[\rho^2,y]\ne 0$.

\end{rem}

\section*{Acknowledgement} I would like to thank Mil\'an Mosonyi for useful
comments on the manuscript. I would also like to thank the anonymous referee for a number of valuable suggestions,
especially for the proof of Lemma \ref{lemma:ctns}. The work was supported by the grants
VEGA 2/0032/09 and meta-QUTE ITMS 26240120022.

\end{document}